\documentclass[11pt]{article}

\begin{document}
\title{\huge{Structure formation by nanosilica particles suspended in levitated droplet}}
\author{Abhishek Saha, Erick Tijerino, Ranganathan Kumar \\
Department of Mechanical Materials and Aerospace Engineering \\
University of Central Florida \\
Orlando, FL 32816 \\ \\
Saptarshi Basu \\ 
Department of Mechanical Engineering \\
Indian Institute of Science \\
Bangalore, India
}

\maketitle

\begin{abstract}
Preferential accumulation and agglomeration kinetics of nanoparticles suspended in an acoustically levitated water droplet under radiative heating has been studied. At high concentrations of nanosilica particles, viscosity of the solution increases, and correspondingly centrifugal motion dominates due to droplet rotation about the levitator axis, which results in the formation of a ring structure. This horizontal ring eventually reorients itself due to an imbalance of acoustic forces on the ring, exposing higher area for laser absorption and subsequent sharp temperature rise.\end{abstract}

\section{Introduction}
\label{Introduction}
Recently, we studied acoustically levitated droplets, containing soluble and insoluble particles under laser irradiation$^{1-3}$. Heating of nanosilica suspended solutions with different initial concentration shows a fascinating agglomeration and structure formation. In the initial stages of heating when significant evaporation has not taken place, the hydrodynamic effects caused by acoustic streaming is stronger compared to other effects such as orthokinetic aggregation. Soon, the evaporate rate becomes stronger resulting in a sharp diameter reduction. At the end of this stage, accumulation of nanosilica sets in as the droplet takes the shape of a bowl due to acoustic pressure difference, and the drop size stops reducing further. As the solvent is depleted, the hydrodynamic effect becomes weaker. This marks the onset of the structure formation stage which is dominated by aggregation or agglomeration of nanosilica particles. The current results show two different structures depending on the initial solute concentration. For nanosilica concentration of less than 1.3\%, the droplet maintains the bowl structure. On the other hand, for concentration greater than 1.9\%, the initial bowl transforms into a horizontal ring. For the concentrations between 1.3 and 1.9, the droplet either forms a ring or a bowl. The formation of ring can be explained as follows. Further increase in viscosity with concentration decreases the strength of recirculation. The centrifugal effect due to droplet rotation about levitator axis becomes stronger than recirculation resulting in the accumulation of particles around the droplet equatorial plane. Thus a horizontal ring is formed due to asymmetries in mass distribution. The horizontal ring first starts oscillating within the acoustic field and eventually reorients itself to form a vertical ring due to an imbalance of forces.

Submitted video is an adapted version of high speed images recorded during heating a 3\% nano-silica droplet. All the stages mentioned earlier can be observed in the video. Initially, it shows a sharp diameter reduction, followed by a formation of a bowl structure. Then the bowl takes shape of a horizontal ring. Eventually this ring reorients itself to become vertical ring.

\vspace{0.5in}

\noindent
{\bf {\large References}}

\noindent
$^1$A Saha, S Basu, C Suryanarayana, R Kumar, "Experimental analysis of thermo-physical processes in acoustically levitated heated droplets", International Journal of Heat and Mass Transfer 2010, Volume 53 (2010),  pp 5663-5674

\noindent
$^2$R Kumar, E Tijerino, A Saha, S Basu, "Structural morphology of acoustically levitated and heated nanosilica droplet", Applied Physics Letters, 97 (2010),  pp 123106

\noindent
$^3$A Saha, S Basu, E Tijerino, R Kumar, "Particle Image Velocimetry and Infrared Thermography 
in a Levitated Droplet with Nanosilica Suspensions", Experiments in Fluid (under review).
\end{document}